\documentclass[%
reprint,
superscriptaddress,
nofootinbib,
amsmath,
amssymb,
aps,
prd,
]{revtex4-2}

\usepackage{graphicx}
\usepackage{dcolumn}
\usepackage{bm}
\usepackage{hyperref}
\usepackage{color}
\usepackage{subfig}
\usepackage{orcidlink}

\begin{document}

\preprint{APS/123-QED}
\renewcommand{\thefootnote}{\fnsymbol{footnote}}
\title{Investigation of Hadronic Cross Sections of Cosmic Ray Carbon and Oxygen on BGO from 200 GeV to 10 TeV energy at the DAMPE Experiment}

\author{F.~Alemanno,\orcidlink{0000-0003-1065-2590}}
\affiliation{Dipartimento di Matematica e Fisica E. De Giorgi, Universit\'{a} del Salento, I-73100, Lecce, Italy}
\affiliation{Istituto Nazionale di Fisica Nucleare (INFN)- Sezione di Lecce, I-73100, Lecce, Italy}
\author{Q.~An}
\altaffiliation[]{Deceased}
\affiliation{State Key Laboratory of Particle Detection and Electronics, University of Science and Technology of China, Hefei 230026, China} 
\affiliation{Department of Modern Physics, University of Science and Technology of China, Hefei 230026, China}
\author{P.~Azzarello}
\affiliation{Department of Nuclear and Particle Physics, University of Geneva, CH-1211 Geneva, Switzerland}
\author{F.~C.~T.~Barbato,\orcidlink{0000-0003-0751-6731}}
\affiliation{Gran Sasso Science Institute (GSSI), Via Iacobucci 2, I-67100 L’Aquila, Italy}
\affiliation{Istituto Nazionale di Fisica Nucleare (INFN)- Laboratori Nazionali del Gran Sasso, I-67100 Assergi, L’Aquila, Italy}
\author{P.~Bernardini,\orcidlink{0000-0002-6530-3227}}
\affiliation{Dipartimento di Matematica e Fisica E. De Giorgi, Universit\'{a} del Salento, I-73100, Lecce, Italy}
\affiliation{Istituto Nazionale di Fisica Nucleare (INFN)- Sezione di Lecce, I-73100, Lecce, Italy}
\author{X.~J.~Bi}
\affiliation{Institute of High Energy Physics, Chinese Academy of Sciences, Yuquan Road 19B, Beijing 100049, China}
\affiliation{University of Chinese Academy of Sciences, Yuquan Road 19A, Beijing 100049, China}
\author{H.~Boutin,\orcidlink{0009-0004-6010-9486}}
\affiliation{Department of Nuclear and Particle Physics, University of Geneva, CH-1211 Geneva, Switzerland}
\author{I.~Cagnoli,\orcidlink{0000-0001-8822-5914}}
\affiliation{Gran Sasso Science Institute (GSSI), Via Iacobucci 2, I-67100 L’Aquila, Italy}
\affiliation{Istituto Nazionale di Fisica Nucleare (INFN)- Laboratori Nazionali del Gran Sasso, I-67100 Assergi, L’Aquila, Italy}
\author{M.~S.~Cai,\orcidlink{0000-0002-9940-3146}}
\affiliation{Key Laboratory of Dark Matter and Space Astronomy, Purple Mountain Observatory, Chinese Academy of Sciences, Nanjing 210023, China}
\affiliation{School of Astronomy and Space Science, University of Science and Technology of China, Hefei 230026, China}
\author{E.~Casilli,\orcidlink{0009-0003-6044-3428}}
\affiliation{Dipartimento di Matematica e Fisica E. De Giorgi, Universit\'{a} del Salento, I-73100, Lecce, Italy}
\affiliation{Istituto Nazionale di Fisica Nucleare (INFN)- Sezione di Lecce, I-73100, Lecce, Italy}
% ...
\collaboration{DAMPE Collaboration}
\altaffiliation{dampe@pmo.ac.cn}
\date{\today}
\begin{abstract}
The Dark Matter Particle Explorer (DAMPE) has made significant progress in measuring the fluxes of cosmic rays. These new measurements are pivotal in advancing our understanding of the origins and propagation mechanisms of cosmic rays. The Bismuth Germanium Oxide (BGO) calorimeter plays a crucial role in these measurements, particularly in the precise determination of cosmic ray fluxes. However, for a calorimetric experiment like DAMPE, uncertainties in hadronic models persist as a major barrier in achieving more accurate measurements of fluxes of cosmic ray nuclei. 
This study centers on the measurement of the inelastic hadronic cross sections of carbon and oxygen nuclei interacting BGO crystals target over an extensive energy range, spanning from 200 GeV to 10 TeV. For carbon nuclei interacting with the BGO target. 
The measurements of the cross sections have achieved a total relative uncertainty of less than 10\% below 8 TeV for carbon, and below 3 TeV for oxygen.
For oxygen nuclei, the same level of precision was attained below 3 TeV. Additionally, we compare the experimental results with Geant4 and FLUKA simulations to validate the accuracy and consistency of these simulation tools. Through comprehensive analysis of the inelastic hadronic interaction cross sections, this research provides validation for the hadronic interaction models used in DAMPE's cosmic-ray flux measurements.
\end{abstract}
\maketitle
\clearpage

\section{Introduction}

The Dark Matter Particle Explorer (DAMPE) \cite{DAMPEmission_1} is a satellite-borne calorimetric detector designed to observe high-energy electrons \cite{electrons_2}, gamma rays \cite{Gamma_22}, and cosmic rays (CRs) \cite{proton_3, He_4, COratios_5,PandHe,DAMPE_Boron}. 
CRs, according to the Fermi acceleration mechanism \cite{Fermi_18}, are expected to follow a single power-law distribution in their energy spectrum, as emitted by astrophysical sources. Galactic CRs are accelerated by sources such as supernova remnants, reaching energies up to a few PeV. During their propagation through interstellar space of the Milky Way, primary CRs.
The current generation of space experiments, including DAMPE \cite{proton_3, He_4, COratios_5,PandHe,DAMPE_Boron}, the Alpha Magnetic Spectrometer Experiment (AMS-02) \cite{AMS_19,AMS_proton,AMS_He,AMS_Li,AMS_C,AMS_Fe,AMS_total} and The CALorimetric Electron Telescope (CALET) \cite{CALET_20,CALET_Proton,CALET_He,CALET_B,CALET_CO}, have advanced the measurement of the nuclei fluxes in CRs , thereby providing valuable insights into the origin and propagation of CRs \cite{BCratio, BCratio2,AMS_theory1,AMS_theory2}.

In case of the DAMPE experiment, the Bismuth Germanium Oxide (BGO) calorimeter is a key sub-detector and plays a pivotal role in these measurements. However, for calorimetric experiments like DAMPE, uncertainties in hadronic models still pose a significant obstacle to obtaining more accurate energy measurements of nuclei. Among these uncertainties, the one stemming from the hadronic interaction cross section is one of the main contributors. 
In the DAMPE-based cosmic ray flux measurements, the maximum systematic uncertainties originating from hadronic interaction models reach 10\% for protons, 15\% for helium, and 20\% for boron at the highest energy \cite{proton_3, He_4,DAMPE_Boron}. Similarly, in calorimetric experiments like CALET, model-induced uncertainties constitute the dominant systematic error source in cosmic ray flux measurements, with maximum uncertainties of 9.2\% and 12.2\% observed in their carbon and oxygen measurements, respectively \cite{CALET_He,CALET_Proton,CALET_B,CALET_CO}.
Consequently, studying the hadronic interaction cross sections of nuclei with detector materials can help validate the calorimeter's response, provide better estimates of systematic uncertainties, and improve the accuracy of CR fluxes.

Currently, ground-based accelerator experiments have measured proton cross sections in the center-of-mass energy range of TeV, corresponding approximately to the PeV kinetic energy range \cite{proton_crosssection, proton_crosssection2}. However, for carbon and oxygen nuclei, measurements are limited to a few GeV per nucleon, with very sparse experimental data available at higher energies \cite{Crosssection_1, Crosssection_2}. Cosmic rays, as a natural particle source for high-energy particle physics experiments, offer a unique opportunity to measure the cross sections of nuclei in the high-energy regime. For instance, AMS-02 data has been used to measure nuclear interaction cross sections for carbon targets, covering a wide rigidity range from GV to TV \cite{AMS_CX_21}. Similarly, the DAMPE collaboration has measured the interaction cross sections of protons and helium nuclei with the BGO target using machine learning methods, with measurements spanning from 20 GeV to 10 TeV \cite{proton_crosssection3}.
DAMPE conducted heavy ion beam tests at CERN to evaluate the energy response of the detector at 40 GeV/n (469 GeV for carbon) and 75 GeV/n (889 GeV for carbon) and compared the results with simulations from Geant4 and FLUKA \cite{EnergyResponse}.
\par
Carbon-12 ($^{12}$C) and oxygen-16 ($^{16}$O) are the dominant isotopes of their respective elements in the universe \cite{COIsotope}, and their high abundance in CRs provides a critical advantage for precise measurement. These two elements play an important role in CR physics, particularly as key probes of CR propagation through their boron-to-carbon (B/C) and boron-to-oxygen (B/O) flux ratios.
In this paper, we present the cross section results for these cosmically abundant nuclei (carbon and oxygen) interacting with BGO material using DAMPE data up to an energy of 10 TeV. 
Measuring their nuclear interaction cross sections not only provides more accurate spectral data, aiding in the in-depth study of cosmic ray origins and propagation, but also validates the precision of DAMPE's detection capabilities for heavier nuclei in the TeV regime.
\par
The paper is organized as follows: Section 2 provides an overview of the DAMPE detector. Section 3 describes the analysis process and presents estimation of systematic uncertainties. Section 4 introduces the results of the hadronic cross section measurements. A concise summary is then given in section 5.

\section{DAMPE detector}
DAMPE is a calorimetric satellite-borne detector that was launched on December 17, 2015. The instrument is structured from top to bottom, including a Plastic Scintillator strip Detector (PSD), a Silicon-Tungsten tracKer (STK), a BGO calorimeter, and a NeUtron Detector (NUD) as shown in Figure 1. 

\textbf{PSD.} The PSD is utilized for identifying the charge of incoming charged particles by measuring their energy loss rate $dE/dx$ \cite{Bethe-Bloch}. It consists of 82 plastic scintillator bars organized into two planes, with the planes oriented orthogonally to each other, and each plane configured in a double-layer arrangement. Each bar measures 88.4 cm $\times$ 2.8 cm $\times$ 1.0 cm in size. With a single-layer efficiency greater than or equal to 0.95, the overall efficiency of the PSD for charged particles exceeds 0.9975 \cite{PSD_7}, providing accurate measurements of particle charge with resolutions of 0.18e for carbon \cite{PSD_ChargeMeasure}.

\textbf{STK.} The STK consists of six layers of silicon micro-strip detectors, each comprising X and Y layers with 64 detectors. The size of each silicon detector is 95 mm $\times$ 95 mm $\times$ 0.32 mm. Additionally, the STK includes three layers of tungsten plates, each with a thickness of 1 mm. 
The total thickness of the STK corresponds to about one radiation length, primarily due to the tungsten layers. The STK offers precise particle track reconstruction with a pitch of 121 $\mu$m and spatial resolution below 80 $\mu$m within the angular acceptance of the detector, i.e., for incidence angles less than or equal to $60^{\circ}$ \cite{STK_8}.

\textbf{BGO.} The BGO calorimeter is responsible for energy measurements and consists of 308 BGO crystal bars organized into 14 layers, with each layer comprising 22 Bi$_4$Ge$_3$O$_{12}$ crystal bars. This configuration provides 32 radiation lengths and 1.6 nuclear interaction lengths. 
The size of each BGO crystal is 25 mm $\times$ 25 mm $\times$ 600 mm, encapsulated in a carbon fiber box with a wall thickness of 0.5 mm.
The BGO calorimeter can measure incident hadron energies from 50 GeV to 500 TeV and has an energy resolution of 40\% at 800 GeV \cite{BGO_9,BGOreadout,BGOrange}.

\textbf{NUD.} The primary function of the NUD is to facilitate electron/hadron identification by utilizing the neutrons generated in hadronic showers initiated within the BGO calorimeter \cite{NUD_10}. The NUD is not applicable to this study and was therefore not used in the analysis.

\begin{figure}[!htb]
	\centering
	\includegraphics[width=3.4in] {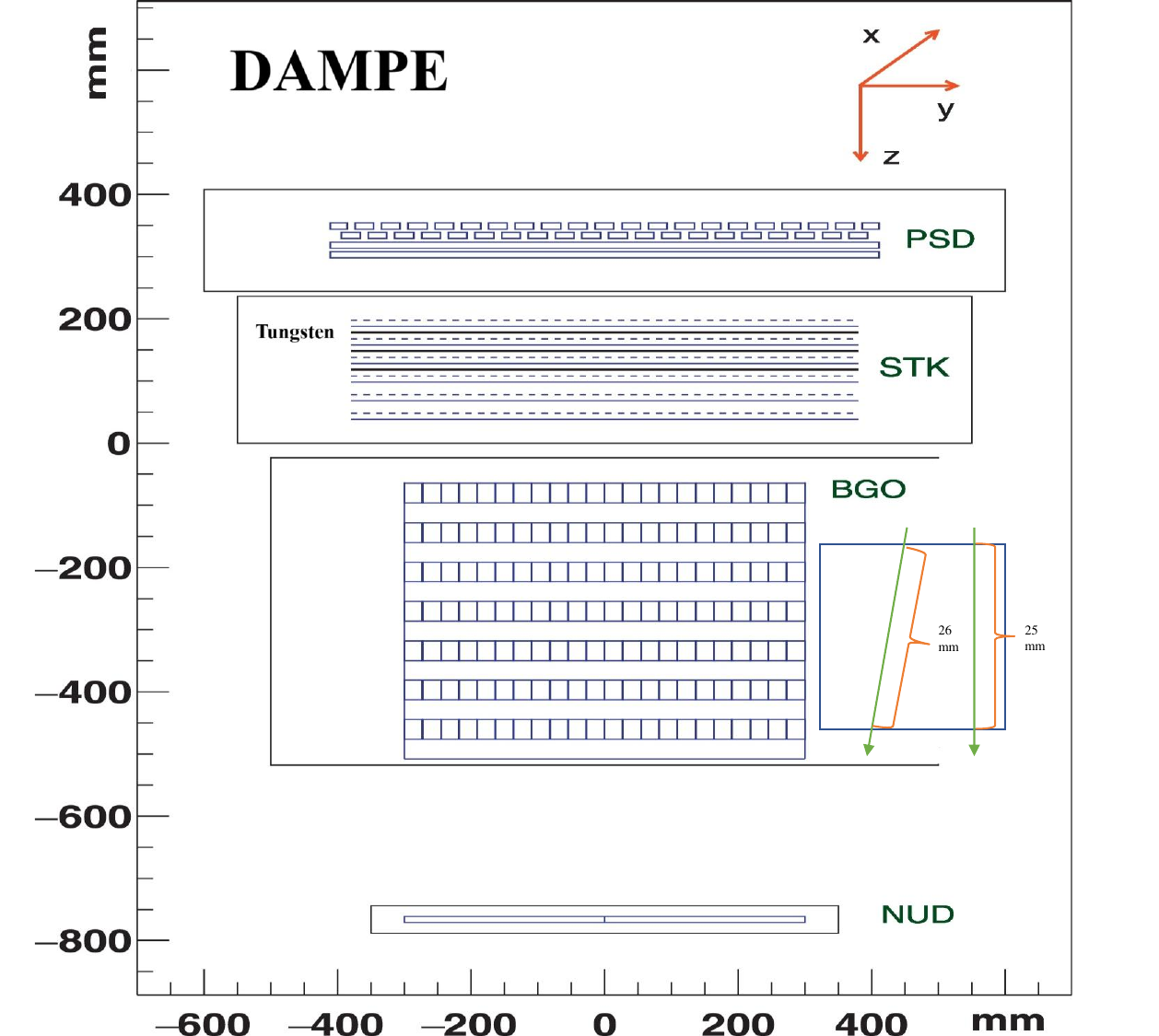}
	\caption{The architecture of DAMPE.}
	\label{fig1}
\end{figure}
\section{Measurement of nuclear interaction cross sections}
Our goal is to obtain hadronic cross sections by measurements of survival probability in the BGO calorimeter. When CR nuclei pass through the BGO material, their survival probability to not experiencing inelastic hadronic interactions is denoted as \( \varepsilon_{\text{sur}} \),
\begin{eqnarray}
	\varepsilon_{sur} = \frac{N_{leave}}{N_{enter}} = \exp(-n \cdot l \cdot \sigma) \label{eq:1}
\end{eqnarray}
where $n$ is the number of target nuclei per unit volume, $l$ is the path length on the target, and $\sigma$ is the interaction cross section on the target. CRs from different directions will pass through different path lengths in the detector, resulting in different survival rates. We use the trajectories reconstructed by the STK detector to distribute and calculate the survival rates for different path lengths.

Different nuclei in CRs are distinguished using the PSD and STK detectors, while the energy of the incident particles is measured using the BGO calorimeter. Due to its limited thickness, approximately 1.6 nuclear interaction lengths, only a portion of the nuclei's incident energy is deposited in the DAMPE calorimeter. We use the unfolding process to recover the kinetic energy of the incident particle from the energy deposited in the calorimeter \cite{Bayestheorem_23}. The relationship between the number of observed events (\( N_{\text{dep}} \)) in the \( i \)-th deposited energy bin and the number of incident events (\( N_{\text{inc}} \)) in the \( j \)-th incident energy bin is expressed as follows:
\begin{eqnarray}
	N_{\text{dep}, i} = \sum_{j} M_{ij} N_{\text{inc}, j}  \label{eq:2}
\end{eqnarray}
where the response matrix \( M_{ij} \) represents the probability that events from the \( j \)-th incident energy bin contribute to the \( i \)-th deposited energy bin.

\subsection{Data Sample}

The Flight Data (FD) analyzed in this study covers the period from January 1, 2016, to December 31, 2023. DAMPE consistently collects about 5 million events per day and has been operating in a stable configuration since its launch \cite{DAMPEdetector_6}. We also evaluated the aging phenomenon of the BGO calorimeter and demonstrated the stability of the BGO calorimeter \cite{BGOstable_11}.
Events recorded while the detector is traversing the South Atlantic Anomaly region are excluded. The primary simulation tool employed in the DAMPE experiment is the Geant4  simulation framework \cite{Geant4_12, Geant4_2, Geant4_3}, version 4.10.5, for conducting Monte Carlo (MC) simulations involving carbon and oxygen nuclei.
For this analysis, the FTFP\_BERT hadronic model from Geant4 is chosen, which spans an energy range from 10 GeV to 100 TeV.
Additionally, MC events in the energy range of 10 GeV to 100 TeV were generated using FLUKA with DPMJET-III \cite{FLUKA_13, DPMJET3}, specifically version 2011.2x.7. 

\textit{\textbf{Preselection}} 

Events are required to pass the high-energy (HE) trigger condition, which demands an energy deposition equivalent to 10 Minimum Ionizing Particles (MIPs, minimum ionizing energy is 23 MeV) in the first three layers and 2 MIPs in the fourth layer of the BGO calorimeter \cite{Trigger}. Cosmic MIP nuclei with \( Z > 4 \) (e.g., Be, B, C, O) depositing ionization energy greater than 16\((Z^2)\) MIPs , where \( Z \) represents the nuclear charge, which satisfies the HE trigger condition, with a trigger efficiency greater than 96\%.

\textit{\textbf{Charge Selection}} 

The ionization energy loss within the PSD is primarily used for reconstructing the charge of each event. As the ionization energy loss of cosmic nuclei is proportional to \( Z^2 \) in the PSD. A comprehensive charge reconstruction algorithm is employed, including path length correction, light attenuation correction, and light yield saturation correction \cite{PSD_ChargeMeasure, PSD_ChargeMeasure_2}. The STK detector provides additional charge identification capabilities. 
For the selection of carbon samples, the PSD charge range is [5.7, 6.4], and the STK signal range is [1500, 2500]. The selection range for oxygen is [7.7, 8.4] and [2500, 3500]. A two-dimensional plot of PSD charge versus STK signal, as shown in Figure 2, is used to select the samples. The STK signal reconstruction utilizes information from all 12 detector layers, achieving a charge resolution of 27\% for carbon and 30\% for oxygen.
MC simulations are employed to generate samples of Beryllium, Boron, Nitrogen, and Oxygen to investigate contamination in the carbon samples from other nuclides. The findings indicate that carbon samples exhibit contamination at a level of 0.19\% when identified by PSD only and 0.08\% when identified by both PSD and STK.
The influence of PSD charge selection was evaluated by varying the charge selection window, resulting in efficiency variations from -5\% to +5\%, and checking the corresponding flux changes. The effect was found to be less than 1\%.

\textit{\textbf{Classification of Events}} 

Using the precise trajectories provided by the STK detector \cite{Kalman}, the path lengths of particles within the BGO calorimeter are measured. The data are divided into three sets of events based on path length intervals of [25,26] mm, [26,27] mm, and [27,28] mm and denoted as $\Delta l_k$. 

\textit{\textbf{Energy Unfolding}} 

This response matrix \( M_{ij} \) is determined through simulations following the power-law CR energy spectrum, with a power-law index of 2.7. The deposited energy bin (i) ranges from 20 GeV to 2.2 TeV, divided into 10 bins, and the kinetic energy bin (j) ranges from 215 GeV to 10 TeV, divided into 9 bins. The unfolding process is performed with four iterations. The incident energy spectrum is reconstructed by solving Equation \eqref{eq:2} using the Bayesian method \cite{Bayestheorem_23}. The statistical uncertainties that arise from the measurement of the energy spectrum (i.e., the deposited energy spectrum of the BGO calorimeter) are evaluated by estimating the statistical uncertainty of each \( N_{\text{dep}, i} \) using a Poisson distribution. The systematic uncertainties caused by the matrix are estimated by varying the spectral index between 2.5 and 3.1 and performing multiple Poisson samplings to statistically reconstruct the fluctuations in the energy spectrum, thereby estimating the systematic uncertainties induced by the response matrix.

\begin{figure}[!htb]
	\centering
	\includegraphics[width=3.4in] {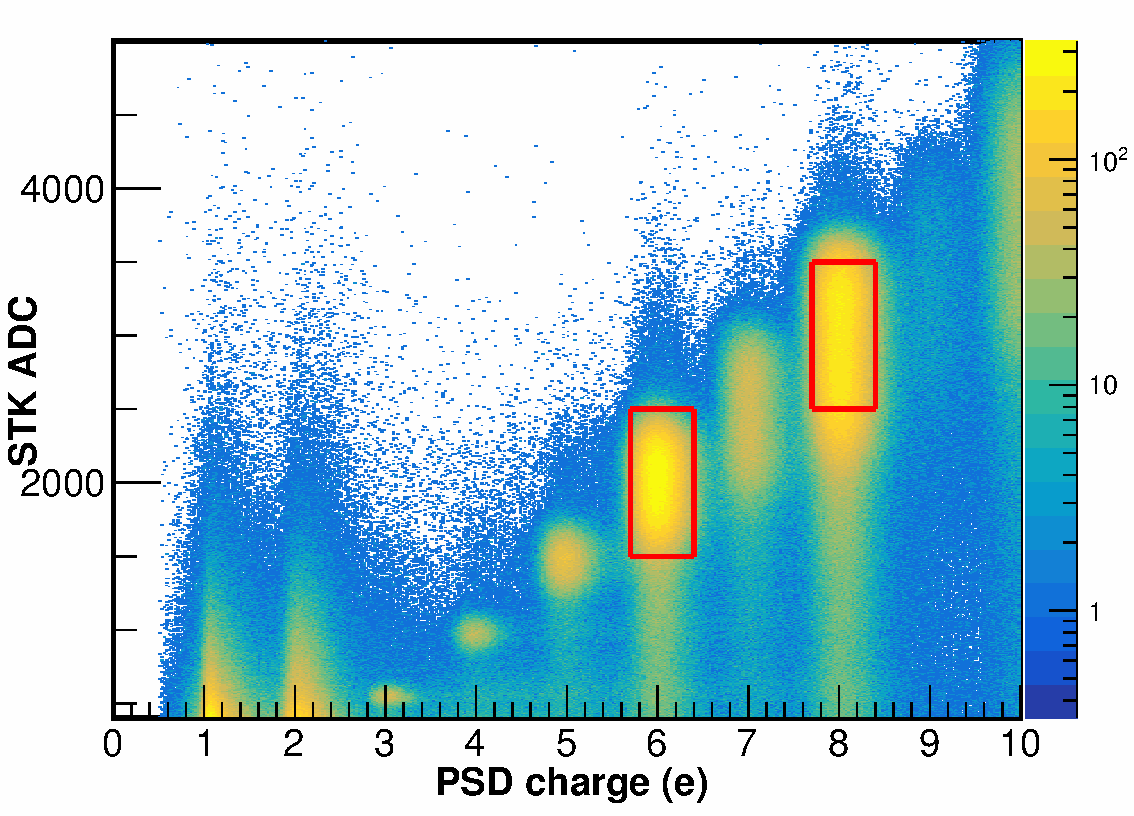}
	\caption{ Distribution plot of the reconstructed charge from PSD and reconstructed signal from STK in FD. The red region represents the charge selection range for carbon and oxygen samples.  }
	\label{fig2}
\end{figure}

\subsection{Measurement of Nuclear Interaction Cross Sections}

\textit{\textbf{Interaction Processes}}  

Cosmic nuclei interact with atoms or molecules in the detector through inelastic processes, primarily involving electromagnetic and hadronic interactions. These interactions generate distinct signals in the detector. For nuclei such as carbon, electromagnetic interactions are predominantly characterized by ionization processes, where the resulting signal is proportional to the square of the charge. In the BGO calorimeter, the ionization energy loss for carbon nuclei is approximately 828 MeV (corresponding to BGO reconstructed charge 6) in a single BGO layer. Conversely, hadronic interactions produce hadronic showers, typically resulting in larger signals than those generated by ionization, as shown in Figure 3.

The survival status of cosmic nuclei can be assessed by monitoring their ionization energy loss, particularly through changes in their charge. A change in charge indicates that the particle has undergone an inelastic hadronic interaction, whereas an unchanged charge suggests that the particle only experienced ionization and survived.
We require the charge measured by the PSD to be consistent with that reconstructed from the BGO calorimeter, particles satisfying this condition are referred to as Minimum Ionizing Nuclei (MIN).
This validation allows us to distinguish non-interacting nuclei from those that have undergone hadronic interactions.

\begin{figure}[!htb]
	\centering
	\includegraphics[width=3.4in] {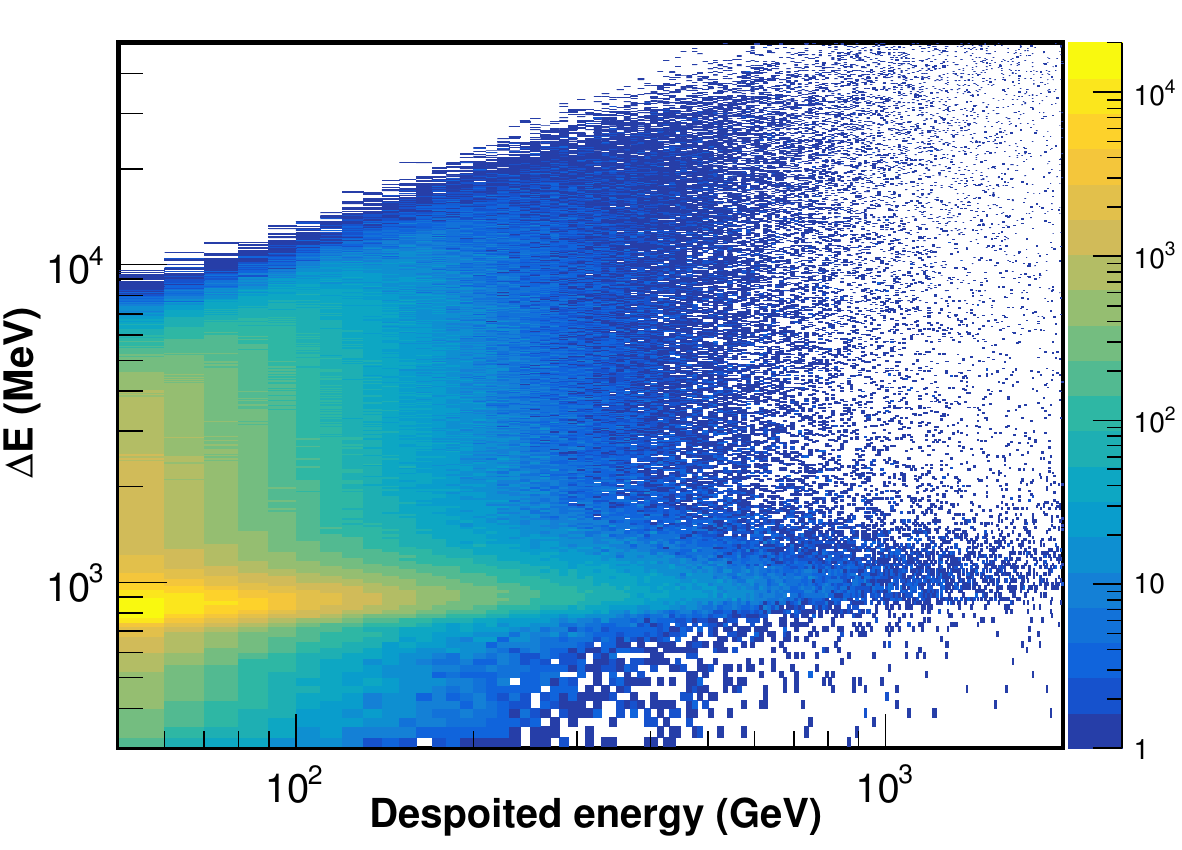}
	\caption{The distribution of the energy loss ($\Delta$ E) with path length correction of carbon samples in the BGO layer as a function of the total deposited energy. }
	\label{fig3}
\end{figure}

\textit{\textbf{Survival probability and cross section}}  

For a given sample of the FD or MC simulation, the event survival is monitored through two layers of the BGO calorimeter. To accurately determine the survival probability of nuclei in the BGO calorimeter, various corrections must be applied to the energy measurements, including those for path length, light attenuation, and light yield saturation \cite{BGO_correction, BGO_correction_2}. The number of events entering the first layer (L1) and having their charge identified is given by \( N_{enter}^{FD} (E^{bgo}_i, \Delta l_k) \), where \( E^{bgo}_i \) is the energy deposited in the BGO as shown in Figure 4(a). After the events pass through the second layer (L2), the number of events leaving and having their charge identified is represented by \( N_{leave}^{FD} (E^{bgo}_i, \Delta l_k) \), as shown in Figure 4(b). 
The charge interval of candidate nuclide (Z) is defined as \(Z - 0.3 < Q_{BGO} < Z + 0.6 \).	
The deposited-energy counts
$N^{\mathrm{FD}}_{\text{enter}}(E^{\mathrm{bgo}}_i,\Delta l_k)$ and
$N^{\mathrm{FD}}_{\text{leave}}(E^{\mathrm{bgo}}_i,\Delta l_k)$ are converted to the incident-kinetic-energy domain with the unfolding procedure (Eq.2), yielding
$N^{\mathrm{FD}}_{\text{enter}}(E^{\mathrm{kin}}_j,\Delta l_k)$ and
$N^{\mathrm{FD}}_{\text{leave}}(E^{\mathrm{kin}}_j,\Delta l_k)$, i.e. the corresponding event counts in each kinetic energy bin $E^{\mathrm{kin}}_j$.
From these quantities, we derive the following relations for survival probability :

\begin{eqnarray}
	\varepsilon_{sur}^{FD} (E^{kin}_j, \Delta l_k) &=& \frac{N_{leave}^{FD} (E^{kin}_j, \Delta l_k)}
	{N_{enter}^{FD} (E^{kin}_j, \Delta l_k)} \nonumber \\
	&=& \exp(-n \cdot l_k \cdot \sigma_{BGO,k}^{FD} (E^{kin}_j)) \label{eq:3}
\end{eqnarray}
\begin{eqnarray}
	\varepsilon_{sur}^{MC} (E^{kin}_j, \Delta l_k) &=& \frac{N_{leave}^{MC} (E^{kin}_j, \Delta l_k)}
	{N_{enter}^{MC} (E^{kin}_j, \Delta l_k)} \nonumber \\
	&=& \exp(-n \cdot l_k \cdot \sigma_{BGO,k}^{MC} (E^{kin}_j)) \label{eq:4} \nonumber \\ 
\end{eqnarray}
where \( n \) [${\rm mm}^{-3}$] is the number of Bi\(_4\)Ge\(_3\)O\(_{12}\)-molecular per volume, and \( \Delta l_k \) [${\rm mm}$] is the path length interval of the nuclei as they traverse the BGO from a given direction.
\(\sigma_{BGO}^{FD} (E^{kin}_j)\) [${\rm mm^2 (barn=10^{-22}mm^2)}$] is the hadronic interaction cross sections in the energy bin-j measured experimentally from the FD-data sets and \(\sigma_{BGO}^{MC} (E^{kin}_j)\) [${\rm mm^2}$] and is the hadronic interaction cross section input by the hadronic model.

\( \sigma_{BGO,k}^{FD}(E^{kin}_j) \) is the hadronic interaction cross section obtained independently from the k-th (k=1,2,3) FD-data sets of nuclide energy bin j using the \eqref{eq:5}
\begin{eqnarray}
	\sigma_{BGO,k}^{FD} (E^{kin}_j) &=& \sigma_{BGO}^{MC} (E^{kin}_j) \nonumber \\ 
	&\cdot& \frac{\ln(\varepsilon_{sur}^{FD} (E^{kin}_j, \Delta l_k))}{\ln(\varepsilon_{sur}^{MC} (E^{kin}_j, \Delta l_k))} \label{eq:5}\nonumber \\
\end{eqnarray}
The survival probability is determined through independent measurements across different energy bins within three path length intervals.

Because k = 1, 2, 3 data sets with different path-length are three independent tests for the same hadronic interaction cross section $\sigma_{BGO}^{FD} (E^{kin}_j)$, the combined survival probability can be constructed and the \eqref{eq:6} and \eqref{eq:7} will be obtained:
\begin{eqnarray}
	\prod \varepsilon_{sur} (E^{kin}_j) &=& \exp \left(-n \left(\sum_k l_k \right) \sigma_{BGO} (E^{kin}_j) \right) \label{eq:6}\nonumber \\
\end{eqnarray}
\begin{eqnarray}
	\sigma_{BGO}^{FD} (E^{kin}_j) &=& \sigma_{BGO}^{MC} (E^{kin}_j) \cdot \frac{\ln \left(\prod \varepsilon_{sur}^{FD} (E^{kin}_j)\right)}{\ln \left(\prod \varepsilon_{sur}^{MC} (E^{kin}_j)\right)} \label{eq:7}
\end{eqnarray}

The hadronic cross section \( \sigma_{BGO}^{FD} (E^{kin}_j) \) obtained from \eqref{eq:7} is comparable with the value averaging over the ones of \(\sigma_{BGO,k}^{FD}(E^{kin}_j)\) (k = 1, 2, 3) obtained using the \eqref{eq:5}. The factors of the \(\sigma_{BGO}^{MC} (E^{kin}_j)\) (\(\sigma_{BGO,k}^{MC}(E^{kin}_j)\)) in the \eqref{eq:5} and \eqref{eq:7} measure the deviations between the hadronic interaction cross sections measured experimentally and the ones assumed by MC-models. The measurements of the hadronic interaction cross sections using FD-MC comparison,
would minimize the impacts of the packing materials(carbon fiber). 

\begin{figure*}[htb]
	\centering
	\subfloat[]{\includegraphics[height=6cm] {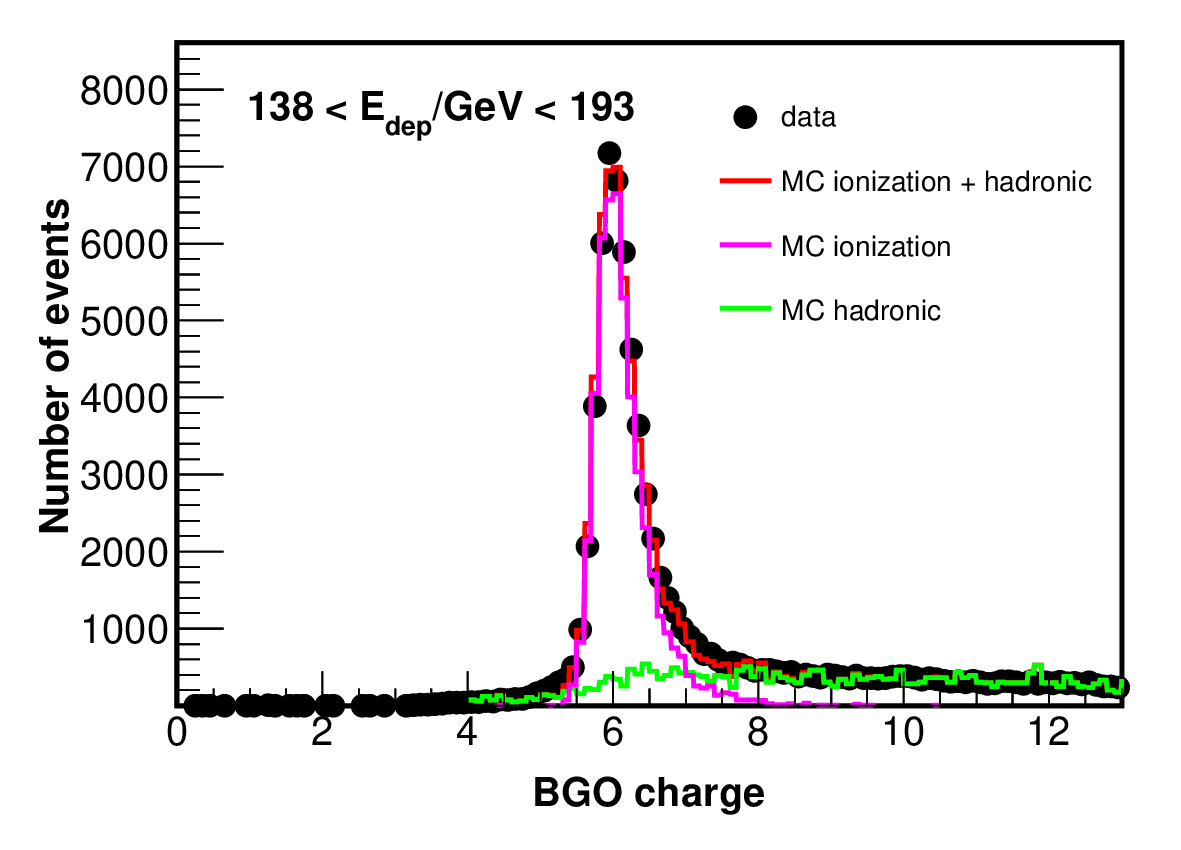}}
	\subfloat[]{\includegraphics[height=6cm] {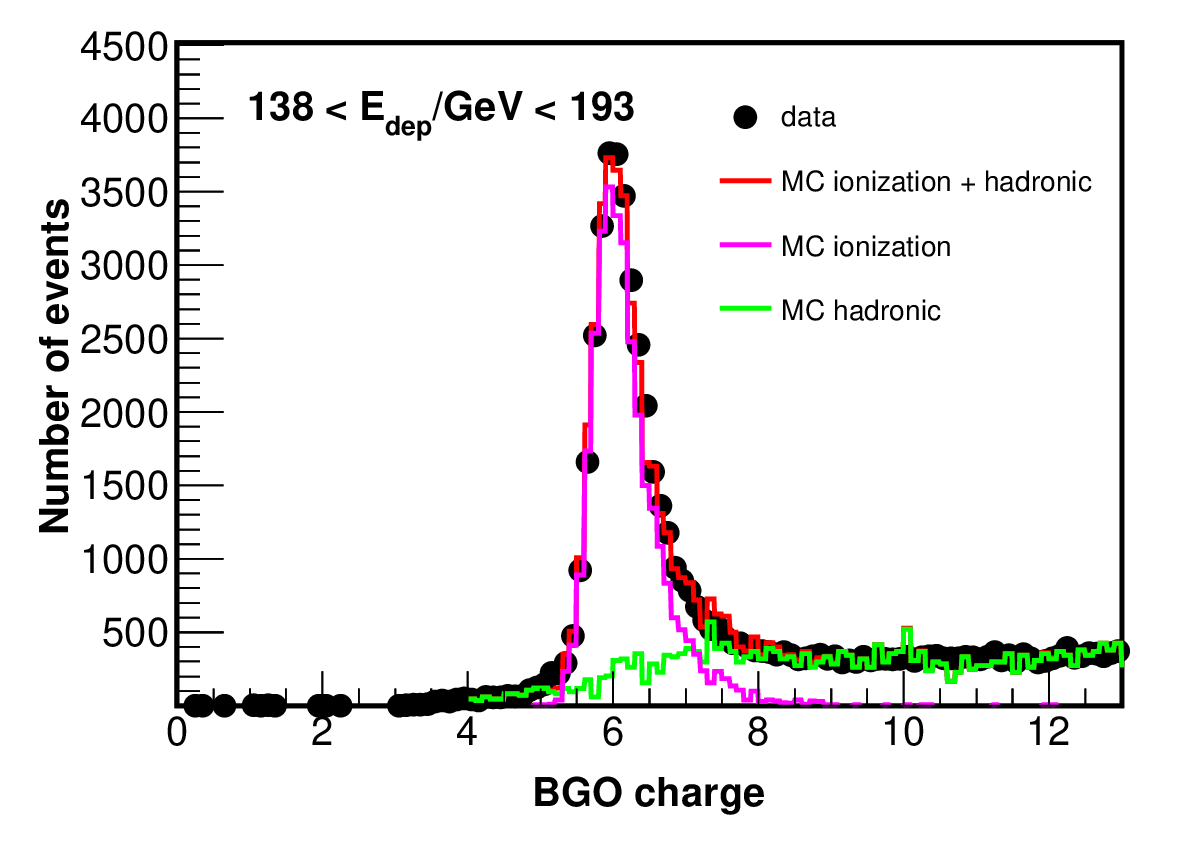}}
	\caption{BGO charge distribution of first layer (a) and second layer (b).}
	\label{fig4}
\end{figure*}

\textit{\textbf{Statistical and systematic uncertainties}}  

The analysis presented herein assesses several sources of systematic uncertainty. Figure 5 provides a comprehensive overview of the energy-dependent systematic uncertainties for carbon.

\textit{\textbf{(i)}} The statistical uncertainty includes not only the statistical error from the survival probability \( \varepsilon_{sur}^{FD} (E^{kin}_j)  \) measured from the FD, but also the statistical error from the survival probability \( \varepsilon_{sur}^{MC} (E^{kin}_j)  \) calculated from the samples generated by the MC simulations. In measurements below 1 TeV, the relative statistical uncertainty from \( \varepsilon_{sur}^{FD} (E^{kin}_j)  \) is approximately 1.1\%, while the relative statistical uncertainty due to the MC-generated sample statistics for \( \varepsilon_{sur}^{MC} (E^{kin}_j)  \) is around 4.4\%. These uncertainties indicate that the statistical error from MC simulations is relatively larger in the lower energy range. Systematic uncertainties associated with the unfolding matrix are considered. Uncertainties from the deconvolution matrix lead to a 1\% uncertainty level.
	
\textit{\textbf{(ii)}} To assess the systematic uncertainty introduced by contamination from other nuclei, we artificially increased the contamination ratio by a factor of 10 to evaluate the impact of the maximum contamination level on the results. The systematic uncertainty introduced by this approach is estimated to be approximately 1.5\%.
		
\textit{\textbf{(iii)}} Systematic uncertainties related to the charge selection range of the BGO detector are evaluated by considering different charge selection ranges, the efficiency for other range variations is maximally 10\% of the efficiency for the baseline. The maximum deviation between the different results is taken as the systematic uncertainty, which is approximately 5\%.
			
\textit{\textbf{(iv)}} 
The differentiation of MIN events refers to the process of distinguishing true non-interacting nuclei from those that undergo hadronic interactions but retain their original charge state within the detector resolution. For instance, events such as \( {}_6^{12} \mathrm{C} + \mathrm{BGO} \rightarrow {}_6^{11} \mathrm{C} + n + \mathrm{BGO} \) (hadronic interactions) might be incorrectly classified as surviving events. Since both $^{12}$C and $^{11}$C carry a charge of six, the current method does not allow for isotope identification between these two carbon species.
MC samples are used to construct two reference templates: one consisting exclusively of non-interacting (purely ionization) events, and the other comprising only hadronic interaction events. These templates are employed in a template fitting procedure applied to the BGO charge distribution, as shown in Figure 4.
The systematic uncertainty is estimated by varying the fraction of the hadronic template in the fit and evaluating the corresponding variations in the reconstructed flux.
The systematic uncertainty from this misclassification is 3\% for energies below 1 TeV, rising to approximately 6\% for energies above 10 TeV.

The uncertainty of the BGO charge selection is estimated by varying the BGO charge selection window to induce efficiency fluctuation ($\pm5\%$) and by checking the flux variations. These checks allow evaluating the systematic uncertainty introduced by the choice of different signal region sizes.

\textit{\textbf{(v)}}
By comparing Geant4 with FLUKA, hadronic model systematic uncertainties can be effectively estimated. The analysis indicates that at lower energies, the uncertainty introduced by varying hadronic interaction models is significant.
	
\begin{figure*}[htb]
	\centering
	\subfloat[]{\includegraphics[height=6cm] {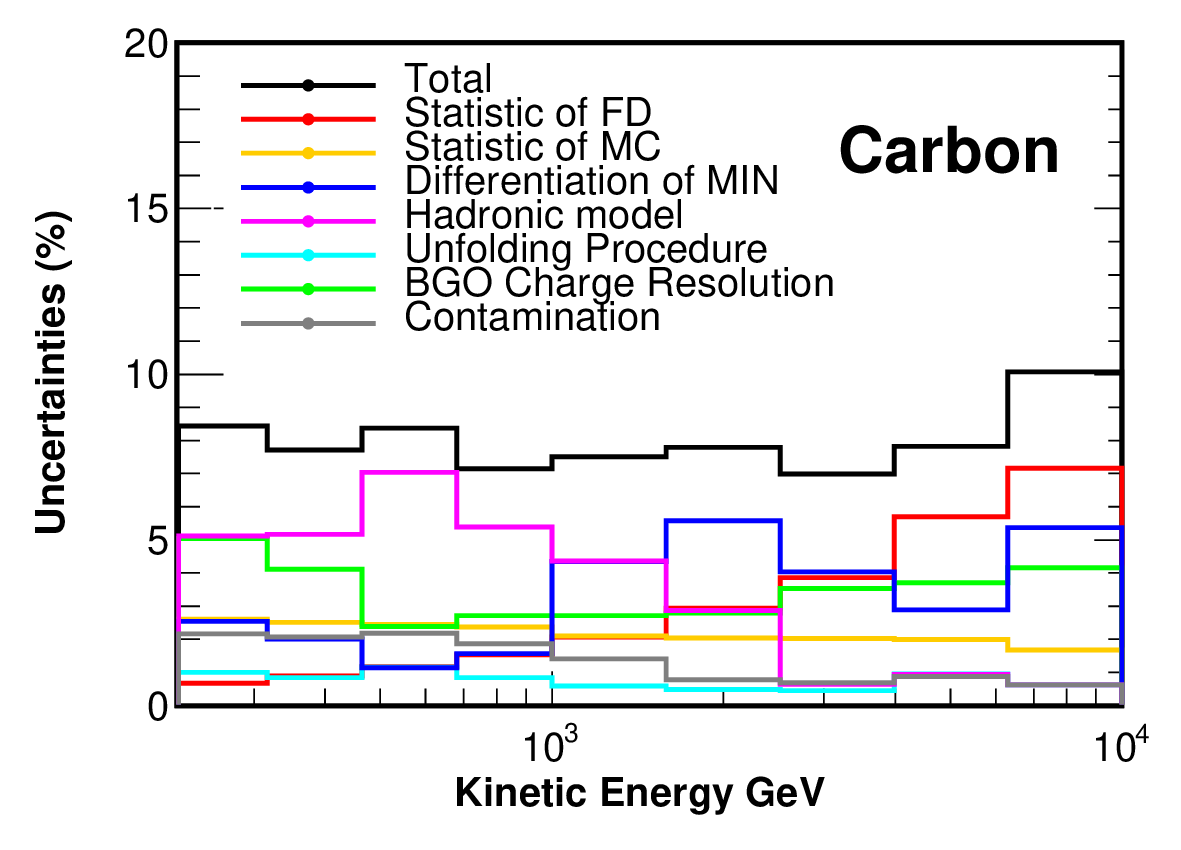}}
	\subfloat[]{\includegraphics[height=6cm] {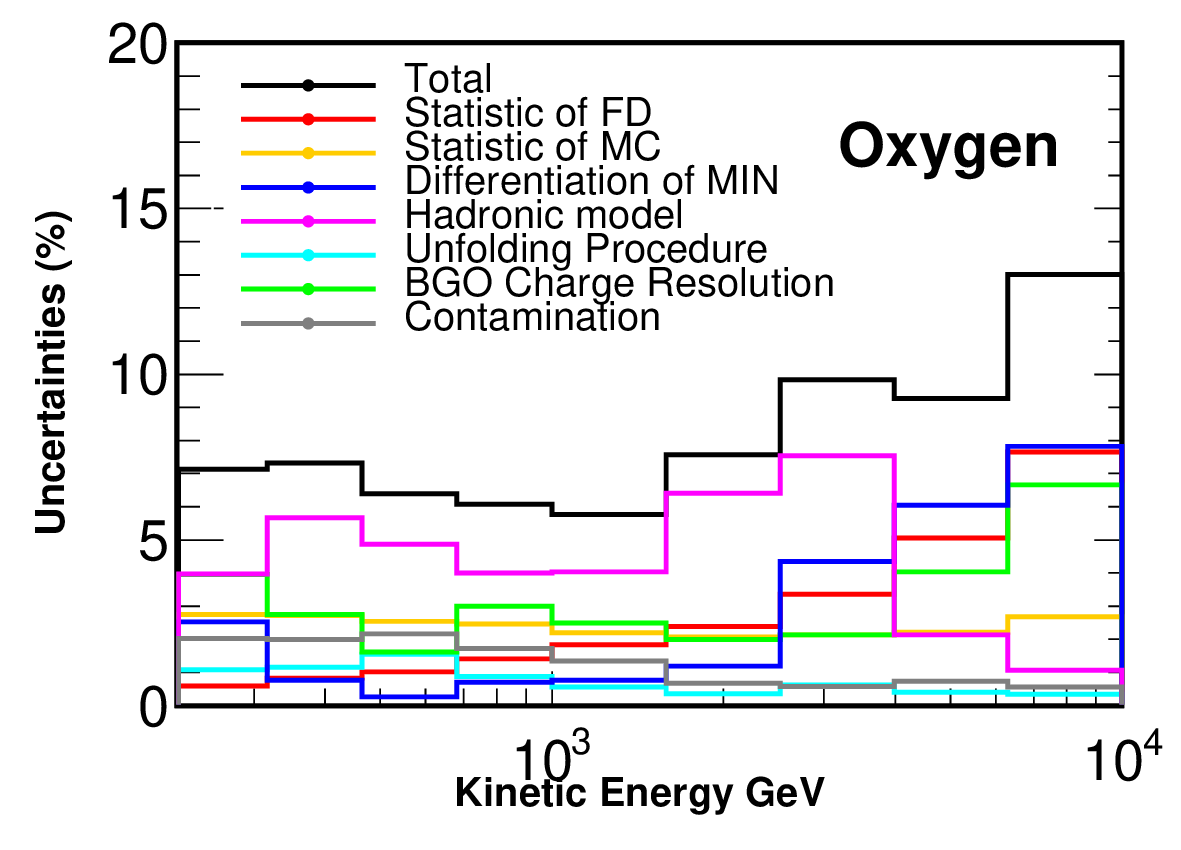}}
	\caption{Illustration of statistical and systematic uncertainties in the measurement of carbon and oxygen inelastic hadronic cross sections. }
	\label{fig7}
\end{figure*}
	
\section{Results}
			

We present the results of various cross section processing methods in Figure 6. In Figure 6(a), we illustrate the cross sections obtained from k = 1, 2, 3 data sets, calculated via Equation \eqref{eq:5}, for different path lengths and energy ranges. It can be observed that the cross sections for different path lengths remain consistent within 1 sigma statistical errors, aligning with theoretical expectations. Figure 6(b) depicts the results obtained using Equation \eqref{eq:7}, where the survival probabilities for different path lengths are combined to calculate the cross sections at different energy (see Table 1). 
The table also presents the inelastic hadronic cross section for a target composed of air, obtained by scaling the cross section ratio from the Geant4 model, as shown in Equation (8). This facilitates their direct application in ground-based cosmic ray observation experiments \cite{Crosssection_3}.

\begin{eqnarray}
	\sigma_{air}^{FD} &=& \sigma_{BGO}^{FD} \cdot \frac{\sigma_{air}^{Geant4}}{\sigma_{BGO}^{Geant4}} \label{eq:8}
\end{eqnarray}
By incorporating all available samples, we significantly diminish statistical uncertainties, leading to more reliable final results. The experimental measurements exhibit excellent agreement with the cross section model employed in the Geant4 simulation software. This model is based on the Glauber-Gribov nucleus-nucleus cross section parameterization \cite{Glaubermodel}. The consistency between experimental data and the model enhances confidence in the reliability of the model.


In a similar manner, the oxygen component is divided into samples, with each evaluated individually to determine the corresponding cross sections. For oxygen, our results provide valuable contributions to the measurement of cross sections on a BGO target, covering the kinetic energy range from 200 GeV to 10 TeV as shown in Figure 7 (see Table 2). 

In this study, we measured the interaction cross sections of high-energy cosmic ray carbon and oxygen with the BGO calorimeter over an energy range of 200 GeV to 10 TeV. Our results indicate that, within the uncertainties, the experimental data agree with the models of the Geant4 and FLUKA, suggesting that these models are reasonably reliable in describing interactions between high-energy CRs and detector materials. This technique holds potential for application to the study of other nucleis in future work, particularly in precision measurements of charge-changing processes.

It is worth noting that the AMS-02 experiment also measured the hadronic interaction cross sections for various nuclei using a carbon target \cite{AMS_CX_21,CERN_CX}. Their measurements exhibit systematic biases compared to the predictions of Geant4. This discrepancy may arise from differences in the types of cosmic ray particles, target material composition, detector structure, or data analysis methods. 

\begin{figure*}[htbp!]
	\centering
	\subfloat[]{\includegraphics[height=6cm] {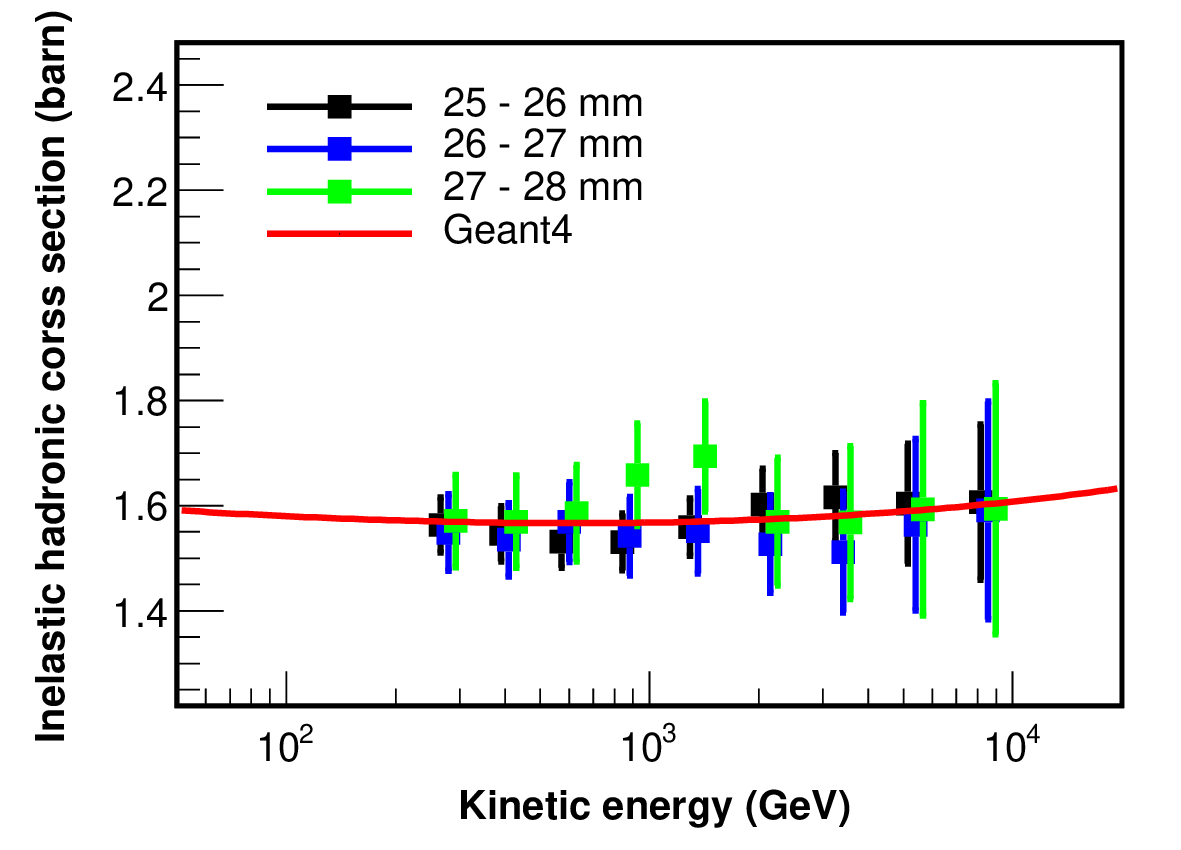}}
	\subfloat[]{\includegraphics[height=6cm] {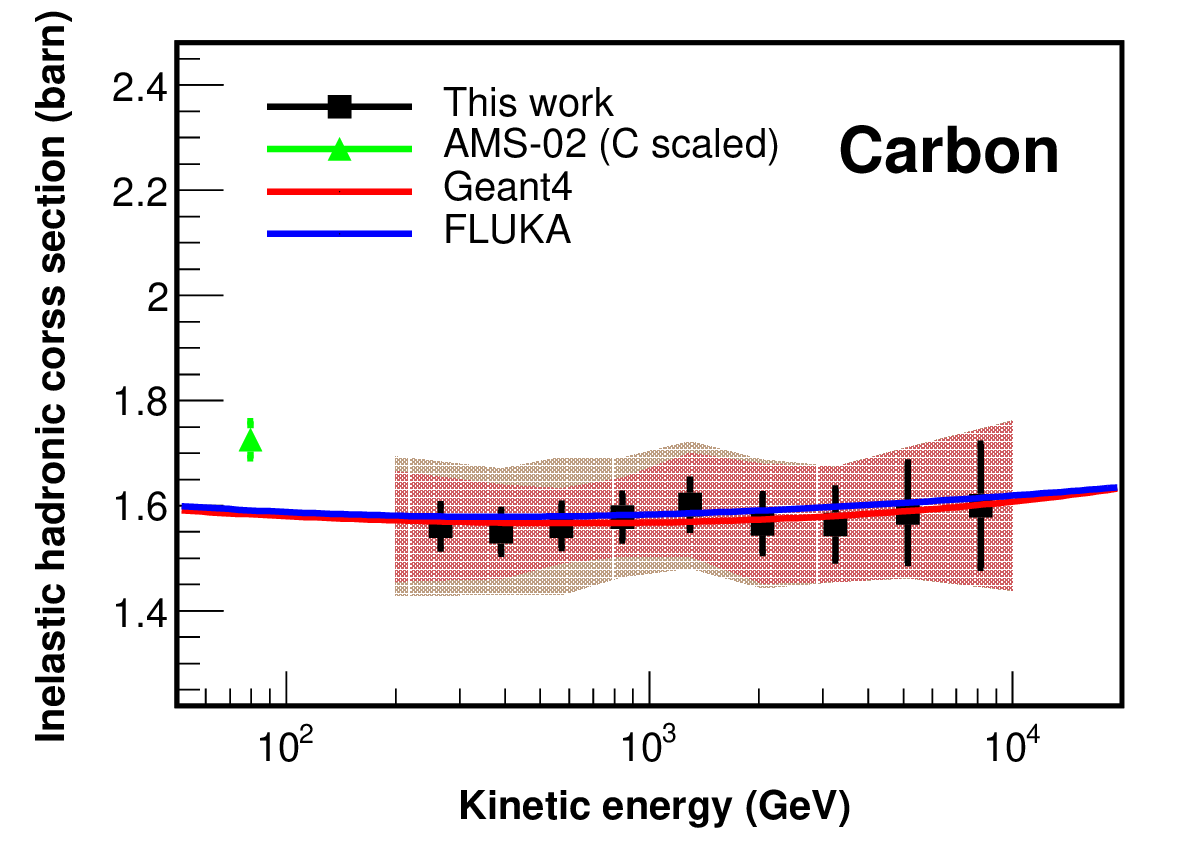}}
	\caption{Inelastic hadronic cross sections of carbon on BGO target with varying path lengths by Eq. \eqref{eq:5} and Eq. \eqref{eq:7}, respectively, shown in figures (a) and (b). The geant4 line indicates the cross sections from Geant4. The blue line represents the cross sections from FLUKA. Error bars represent statistical uncertainties, and the shaded band displays total systematic uncertainties. Inner shaded band shows analysis procedure uncertainties without model uncertainties.
	In figure (b), other measurement for carbon targets (scaled) \cite{AMS_CX_21} are shown for comparison.}
	\label{fig5}
\end{figure*}

\begin{figure}[htbp!]
	\centering
	\includegraphics[width=3.4in] {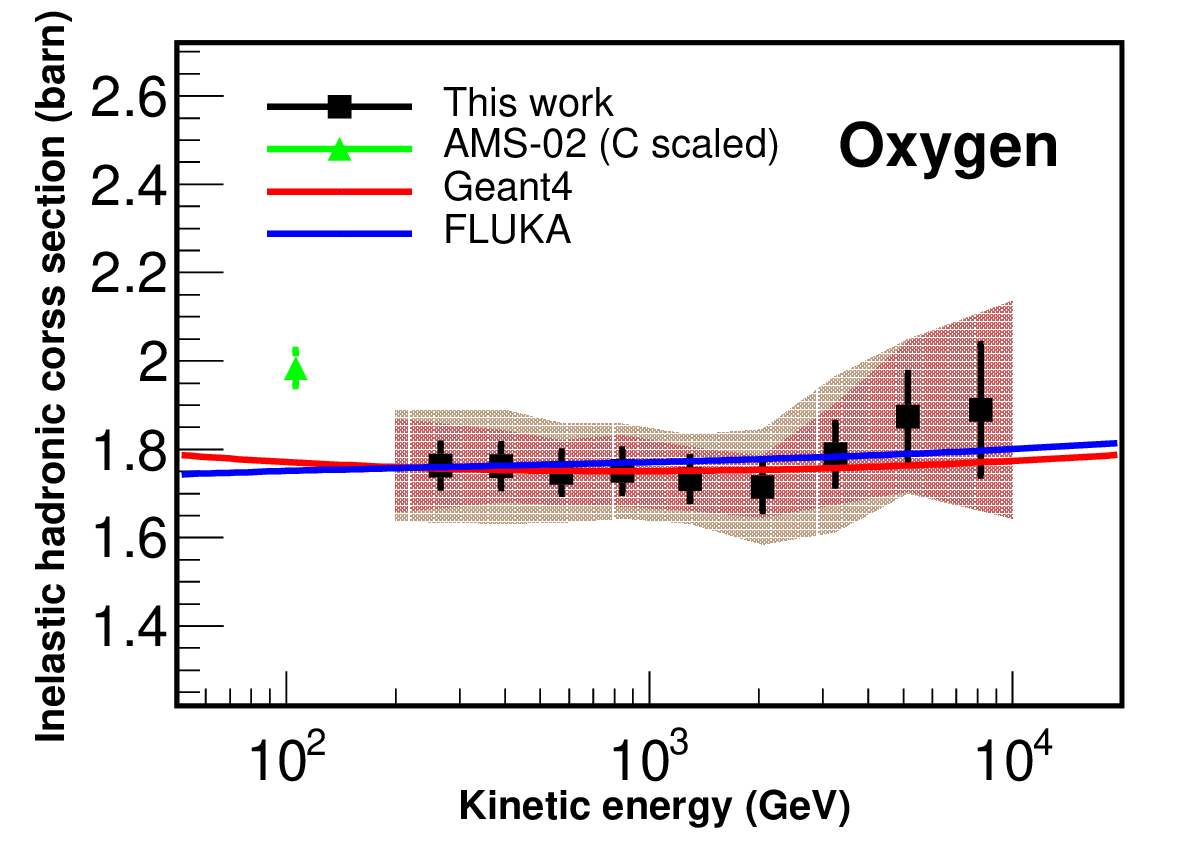}
	\caption{Inelastic cross sections of oxygen on BGO target with Eq. \eqref{eq:7}. The red line represents the cross sections from Geant4. The blue line represents the cross sections from FLUKA. Error bars represent statistical uncertainties, and the outer shaded band displays total systematic uncertainties. Inner shaded band shows analysis procedure uncertainties without model uncertainties.
	In figure, other measurement for carbon targets (scaled) \cite{AMS_CX_21} are shown for comparsion. }
	\label{fig6}
\end{figure}

\begin{table*}[htbp!]
	\centering
	\setlength{\tabcolsep}{8pt}
	\caption{Inelastic hadronic cross sections of carbon on BGO and air targets (scaled). These uncertainties include the statistical uncertainty from the FD, the statistical uncertainty from the MC, the systematic uncertainties related to the analysis (such as BGO charge resolution and MIN differentiation), and finally, the systematic uncertainty arising from the hadronic interaction models.	
	}
	\begin{tabular}{ccc}
		\hline
		\hline		
		\textbf{Energy } & \textbf{  $\sigma_{bgo} \pm stat(FD) \pm stat(MC) \pm sys(Analysis) \pm sys(Model)$ } & \textbf{  $\sigma_{air} \pm stat \pm sys$ } \\
		\textbf{(GeV)} & \textbf{(barn)} & \textbf{(barn)} \\
		\hline		
		265.84 & $1.56 \pm 0.01 \pm 0.04 \pm 0.10 \pm 0.08 $  & $0.96 \pm 0.03 \pm 0.08 $ \\
		390.19 & $1.55 \pm 0.01 \pm 0.04 \pm 0.08 \pm 0.08 $  & $0.96 \pm 0.03 \pm 0.07 $ \\
		572.73 & $1.56 \pm 0.02 \pm 0.04 \pm 0.06 \pm 0.11 $  & $0.96 \pm 0.03 \pm 0.08 $ \\
		840.65 & $1.58 \pm 0.02 \pm 0.04 \pm 0.06 \pm 0.09 $  & $0.97 \pm 0.03 \pm 0.06 $ \\
		1292.45 & $1.60 \pm 0.03 \pm 0.03 \pm 0.09 \pm 0.07 $  & $0.99 \pm 0.03 \pm 0.07 $ \\
		2048.39 & $1.57 \pm 0.05 \pm 0.03 \pm 0.10 \pm 0.04 $  & $0.97 \pm 0.03 \pm 0.07 $ \\
		3246.48 & $1.56 \pm 0.06 \pm 0.03 \pm 0.08 \pm 0.01 $  & $0.97 \pm 0.04 \pm 0.05 $ \\
		5145.32 & $1.59 \pm 0.09 \pm 0.03 \pm 0.08 \pm 0.01 $  & $0.98 \pm 0.06 \pm 0.05 $ \\
		8154.79 & $1.60 \pm 0.11 \pm 0.03 \pm 0.11 \pm 0.01 $  & $0.99 \pm 0.07 \pm 0.07 $ \\		
		\hline
		\hline
	\end{tabular}
	\label{Inelastic cross sections of carbon on BGO target}
\end{table*}

\begin{table*}[htbp!]
	\centering
	\setlength{\tabcolsep}{8pt}
	\caption{Inelastic hadronic cross sections of oxygen on BGO and air targets (scaled). These uncertainties include the statistical uncertainty from the FD, the statistical uncertainty from the MC, the systematic uncertainties related to the analysis (such as BGO charge resolution and MIN differentiation), and finally, the systematic uncertainty arising from the hadronic interaction models.	}
	\begin{tabular}{ccc}
		\hline
		\hline		
		\textbf{Energy } & \textbf{  $\sigma_{bgo} \pm stat(FD) \pm stat(MC) \pm sys(Analysis) \pm sys(Model)$ } & \textbf{  $\sigma_{air} \pm stat \pm sys$ } \\
		\textbf{(GeV)} & \textbf{(barn)} & \textbf{(barn)} \\
		\hline
		265.84 & $1.76 \pm 0.01 \pm 0.05 \pm 0.09 \pm 0.07 $  & $1.12 \pm 0.03 \pm 0.07 $ \\
		390.19 & $1.76 \pm 0.01 \pm 0.05 \pm 0.06 \pm 0.10 $  & $1.12 \pm 0.03 \pm 0.08 $ \\
		572.73 & $1.75 \pm 0.02 \pm 0.04 \pm 0.05 \pm 0.09 $  & $1.11 \pm 0.03 \pm 0.06 $ \\
		840.65 & $1.75 \pm 0.02 \pm 0.04 \pm 0.06 \pm 0.07 $  & $1.11 \pm 0.03 \pm 0.06 $ \\
		1292.45 & $1.73 \pm 0.03 \pm 0.04 \pm 0.05 \pm 0.07 $  & $1.10 \pm 0.03 \pm 0.06 $ \\
		2048.39 & $1.72 \pm 0.04 \pm 0.04 \pm 0.04 \pm 0.11 $  & $1.09 \pm 0.03 \pm 0.07 $ \\
		3246.48 & $1.79 \pm 0.06 \pm 0.04 \pm 0.09 \pm 0.14 $  & $1.14 \pm 0.04 \pm 0.10 $ \\
		5145.32 & $1.87 \pm 0.09 \pm 0.04 \pm 0.14 \pm 0.04 $  & $1.19 \pm 0.06 \pm 0.09 $ \\
		8154.79 & $1.89 \pm 0.14 \pm 0.05 \pm 0.19 \pm 0.02 $  & $1.20 \pm 0.09 \pm 0.12 $ \\
		\hline
		\hline
	\end{tabular}
	\label{Inelastic cross sections of oxygen on BGO target}
\end{table*}

\section{Summary}
This investigation focuses on the inelastic hadronic interaction cross sections in BGO using high-energy CR data collected by DAMPE. The results reveal that the measured inelastic hadronic cross sections for carbon and oxygen nuclei are in excellent agreement with the Geant4 and FLUKA model used in the simulation software, across the energy range from 200 GeV to 10 TeV. 
By leveraging CRs as a “natural accelerator,” this study successfully achieves measurements of the inelastic hadronic interaction cross sections of high-energy carbon and oxygen nuclei on a BGO target in a space-based calorimetric detector. This work also provides a solid experimental foundation for high-precision assessments of cosmic ray fluxes.

\section{Acknowledgment}
The DAMPE mission was funded by the strategic priority science and technology projects in space science of Chinese Academy of Sciences. In China the data analysis is supported by the National Key Research and Development Program of China (No. 2022YFF0503303), the National Natural Science Foundation of China (Nos. 12220101003, 12275266, 12003076, 12022503 and 12103094), Outstanding Youth Science Foundation of NSFC (No. 12022503), the Project for Young Scientists in Basic Research of the Chinese Academy of Sciences (No. YSBR-061), the Strategic Priority Program on Space Science of Chinese Academy of Sciences (No. E02212A02S),the Youth Innovation Promotion Association of CAS (No. 2021450), the Young Elite Scientists Sponsorship Program by CAST (No. YESS20220197), the New Cornerstone Science Foundation through the XPLORER PRIZE and the Program for Innovative Talents and Entrepreneur in Jiangsu. In Europe the activities and data analysis are supported by the Swiss National Science Foundation (SNSF), Switzerland, the National Institute for Nuclear Physics (INFN), Italy, and the European Research Council (ERC) under the European Union’s Horizon 2020 research and innovation programme (No. 851103).

\nocite{*}

\bibliography{mybibfile}
\end{document}